\documentclass[aps,prl,reprint,amssymb,amsmath]{revtex4-1}

\usepackage{graphicx}
\usepackage{algorithm}
\usepackage[noend]{algpseudocode}

\begin{document}

\title{Using path integrals for the propagation of light in a scattering dominated medium}
\author{Gabriel H. Collin}
\email{gabrielc@mit.edu}
\affiliation{Massachusetts Institute of Technology, Cambridge, MA 02139, USA}

\begin{abstract}
The high computational expense of simulating light through ray-tracing in large, sparsely instrumented particle detectors such as IceCube and Antares is a critical outstanding problem in particle physics.  
When the detector is sparsely instrumented, ray tracing is inefficient, as nearly all of these rays are either lost in the bulk of the detector due to absorption or simply fail to end on a detector.      
Particle astrophysics experiments face a similar problem when they simulate cosmic ray muon fluxes in their detectors.    
Many fields of science face calculations that involve constrained initial and final states, with stochastic processes between.     
Taking the case of ray-tracing of light as our example, this paper describes a new and highly computationally efficient approach to the problem.   
By specifying the problem as a path integral, the final state of these rays can be constrained to land on a light sensitive element. 
The path integral can then be efficiently sampled using Reversible Jump Markov Chain Monte-Carlo, yielding performance improvements of up to 1,000 times faster on a realistic test scenario.
\end{abstract}

\maketitle

Particle detectors do not measure the properties of particles directly, they instead measure the quantity of electric charge or light deposited by those particles in the detector.
Inferring the properties of the particles requires calibration.
Sometimes this can be done using a test beam of particles, but often this is not possible or practical.
This task then falls to simulation, where the behavior of the particle detector is modeled ab initio.

In large volume neutrino detectors, the passage of charged particles produces Cherenkov radiation or scintillation light.
This radiation travels through the detector bulk, suffering scattering and absorption before landing on a light sensitive element.
Thus, the scattering and absorption properties of the bulk must be known to accurately simulate this light propagation.
They are inferred in two main steps: first, a known quantity of light is injected into the detector using an artificial light source; second, this process is repeated inside the simulation, and characteristics of the bulk are found when the outputs of both processes match.

The simulation of this light is traditionally performed using a ray tracer.
The ray tracer keeps track of the current position and direction of a photon, and at each iteration, moves it forward and changes the direction according to the scattering model of the bulk.
However, in gigaton scale neutrino detectors -- such as IceCube \cite{AhrensSensitivityIceCubedetector2004,AartsenIceCubeGen2VisionFuture2017} and Antares \cite{AgeronANTARESfirstundersea2011} -- the solid angle to the nearest light sensitive element can be as low as $\mathcal{O}(10^{-6})$.
Thus, many simulated photons get lost in the bulk, terminating their tracing without reaching a light sensitive element.
This leads to a large inefficiency in simulation, and inferring the characteristics of the bulk can require up to 10,000 GPU hours \cite{dimapri}.

This problem is best characterized by a highly constrained initial and final state for the light rays.
Ray tracers perform poorly because they can constrain either the initial or final state, but not both.
To impose both constraints, the entire history of the light ray must be specified up front, forming a path.
The intensity of the light that reaches the light sensitive elements can then be found by solving a path integral.

The use of path integration for the simulation of light propagation was first developed for the rendering of computer generated images, where it is known as Metropolis Light Transport \cite{VeachMetropolisLightTransport1997}.
The scenes that MLT was developed to render are dominated by reflections off hard surfaces, and scattering is usually treated as a perturbation \cite{PremozePathIntegrationLight2003,PremozePracticalRenderingMultiple2004}.
This paper presents the formulation required for the simulation of light in the bulk of neutrino detectors, where the dynamics are dominated by scattering.

\section{Numerical path integration}

Path integration can be approached numerically by specifying the path, $\mathbf{f}$, as a series of straight line segments connected by common vertices, $\vec{f}_n$:
\begin{equation}
	\int_{\Omega} e^{-S[\mathbf{f}]} D\mathbf{f} \approx \int_{\Omega} p(\vec{f}_0, \vec{f}_1, \ldots) D\mathbf{f}.
\end{equation}
Each vertex -- except the first and last -- is a location where the light ray scatters and changes direction.
Instead of deriving the action ($S[\mathbf{f}]$) explicitly, the problem is framed in terms of a probability distribution $p(\ldots)$.
Thus the integral can be sampled using Markov Chain Monte-Carlo tools, which were developed for statistical inference.

The set of all paths, $\Omega$, is a union of multidimensional vector spaces, as the number of scattering points is a variable quantity.
For example, when the scattering length is long, the most probable path will involve few scattering vertices, while a shorter scattering length requires more vertices.
The sampling algorithm must be able to change dimensionality, so as to adapt the number of vertices to the problem at hand.

\begin{figure}
\includegraphics[page=1]{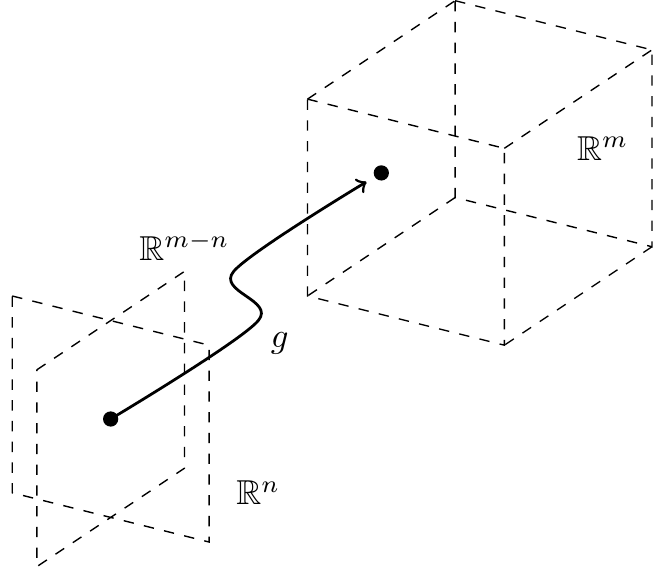}
\caption{\label{fig:gfunc} A point in the product space of $\mathbb{R}^n$ and $\mathbb{R}^{m-n}$ is mapped -- by $g$ -- to a point in the space $\mathbb{R}^m$.}
\end{figure}

Trans-dimensional sampling can be performed using an algorithm called Reversible Jump Markov Chain Monte-Carlo \cite{GreenReversiblejumpMarkov1995}.
This introduces a new proposal stage to the familiar MCMC algorithm.
While a probability distribution $p_n: \mathbb{R}^n \rightarrow \mathbb{R}$ cannot be directly compared to a probability distribution $p_m: \mathbb{R}^m \rightarrow \mathbb{R}$ when $n \neq m$, in the case that $m > n$, $p_n$ can be augmented with an additional distribution $q: \mathbb{R}^{m-n} \rightarrow \mathbb{R}$ so that $p_n \times q$ has the same dimensionality as $p_m$.
After sampling, the extra dimensions introduced by $q$ can be marginalized out to recover the original distribution $p_n$.

These two coordinate spaces are connected by a coordinate transformation function $g: \mathbb{R}^n \otimes \mathbb{R}^{m-n} \rightarrow \mathbb{R}^m$ as shown in Figure~\ref{fig:gfunc}.
This function is an arbitrary choice, but efficient proposals require areas of high probability density in $p_n \times q$ to map to areas of high probability density in $p_m$.

The jump proposal proceeds by drawing a random sample, $\vec{z}$, from $q$.
Using $\vec{z}$ and the current position of the chain $\vec{x}$, the proposal location in the higher dimensional coordinate system is computed from $g$.
The acceptance factor of this proposal is given by
\begin{equation}
	A = \frac{p_n(\vec{x})}{p_m(\vec{y})} \frac{p(m \rightarrow n)}{p(n \rightarrow m)} \frac{1}{q(\vec{z})} \left| \frac{\partial g(\vec{x}, \vec{z})}{\partial \vec{y}} \right|, \label{acceptFact}
\end{equation}
where $|\partial g/\partial \vec{y}|$ is the Jacobian factor for the coordinate transform $g$, and $p(m \rightarrow n)$ and $p(n \rightarrow m)$ are the probabilities of proposing a move from $m$ to $n$ dimensions and vice-versa.
The jump proposal is then accepted with probability
\begin{equation}
	\alpha = \min(1, A).
\end{equation}

The reverse move of reducing the number is dimensions is performed by deterministically computing the proposal coordinates $\vec{x}$ and $\vec{z}$ from the current position of the chain $\vec{y}$ via the inverse coordinate transform $g^{-1}$.
The acceptance factor is computed again according to equation~\ref{acceptFact}, and the jump is accepted with probability
\begin{equation}
	\alpha = \min(1, A^{-1}).
\end{equation}

\section{The path probability distribution}
The path probability distribution is comprised of three main factors: a factor for the probability of emission from the light source, a probability for detection at the photo-sensitive element, and a factor for the probability of light scattering at each intermediate vertex. 
A complete construction of the distribution is detailed in the appendix.

The initial factor is a probability density over the outgoing direction of the light leaving the fixed point source.
The final factor is the cumulative probability of the light reaching the detector without scattering, and the probability of detection conditioned on the point where the path terminates on the detector.

The intermediate factors are the probability density of the light scattering at the intermediate vertex location multiplied by the angular scattering distribution.
This distribution is a probability density over the change of direction that the light undergoes at the vertex.

\subsection{Jump proposal}
Only jumps that add or remove a single vertex are considered, and are often called ``birth/death'' moves \cite{FanReversiblejumpMarkov2010,RichardsonBayesianAnalysisMixtures1997}.
A new vertex is added to the path by placing it between two already existing adjacent vertices with probability proportional to the scattering depth ($\tau_b(k)$ defined in equation~\ref{eq:b_depth_def}) between them.

A vertex is removed by reconnecting its adjacent vertices.
The initial and final vertices cannot be removed, and intermediate vertices are chosen with equal probability.

\begin{figure}
\includegraphics[page=2]{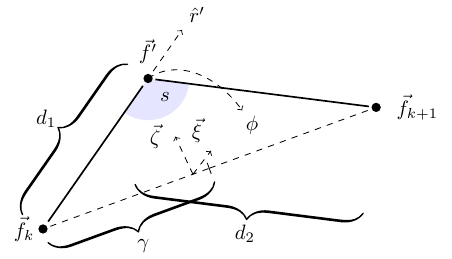}
\caption{\label{fig:bisph} A new vertex $\vec{f}'$ is inserted using a bi-spherical coordinate system $s$, $\phi$ and $t = \ln{(d_1/d_2)}$.}
\end{figure}

The new vertex is placed at a position governed by the distribution $q$.
In practice, the angular scattering distribution is highly forward peaked.
Thus, it is most natural to let $q$ be a distribution over bi-spherical coordinates $s,t,\phi$, as the angle of scattering is directly related to $s$.
Figure~\ref{fig:bisph} shows that the new vertex is placed between the existing vertices, which form the foci of the coordinate system.

\subsection{Path proposal distribution}

The choice of coordinate system for specifying the path has a large effect on the efficiency of the sampler.
In a spherical coordinate system the location of a vertex depends on the coordinates of all vertices before it.
This introduces a large degree of correlation, as small changes in the first few vertices have a lever arm effect on the last vertices. 
The use of Cartesian coordinates removes this effect, but introduces a range of length scales that requires a highly dynamic proposal distribution.

Bi-spherical coordinates are naturally specified in dimensionless quantities, as the length scale is defined by the distance between the foci.
The position of the vertex with index $n/2$ is specified with the start and end points of the path as foci.
The remainder of the path is then specified in a recursive fashion: the path is split into two sets around the current vertex.
The mid point vertex of each set is then specified using the start and end vertices of the set as foci.
The set is then divided in two around the mid point, and the process repeated, creating a tree of coordinates as shown in Figure~\ref{fig:tree_coords}.

To propose a new sample location, the end point of the path on the surface of the detector is first updated using a von Mises-Fisher distribution.
The bi-spherical coordinates of $\theta, t$ are updated using normal distributions.
The bi-spherical coordinate $s$ can induce large changes in the vertex position when $s$ is small, so it is first transformed to $\zeta = \tanh^{-1}{(\cos s)}$ where it is updated with a normal distribution.
The variance of these normal distributions are scaled to be inversely proportional to the number of vertices. 
With 10\% probability, a jump proposal is made instead of updating the current path.
When a jump is made, movements up or down in dimension are chosen with equal probability.

\section{Implementation}

\begin{figure}
\includegraphics[page=3]{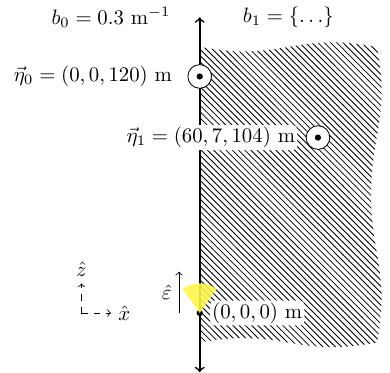}
\caption{\label{fig:setup} The synthetic test case on which the path sampler and ray tracer were tested. The white circles with central points represent the two detectors and their locations. The point with a yellow emission indicates the light source. The preferred direction of both detectors, $\hat{\rho}$ points into the page (negative $y$-direction). The absorption parameter $a$, is $0.01$ m${}^{-1}$ over the entire volume. The scattering parameter $b_1$ ranges over the values listed in the main text.}
\end{figure}

A synthetic test case was constructed to compare the path sampling method against a ray tracer.
The path sampler and ray tracer were both written in \texttt{C++}, targeting the CPU.
The ray tracer is explained in more detail in the appendix.

The angular emission distribution (equation~\ref{eq:emission_dist}) was chosen to be a von Mises-Fisher distribution aligned with the $z$-axis with a concentration parameter, $\kappa$, of 38.
The angular scattering distribution was chosen to be a mixture of a simplified Liu and Henyey-Greenstein distributions (equation~\ref{eq:angular_scattering}).
This mixture approximates the Mie scattering caused by dust impurities in the bulk material \cite{AartsenMeasurementSouthPole2013}.
The bulk was divided into two regions by a plane in the $y$-$z$ axis, as shown in Figure~\ref{fig:setup}.
The region in the negative $x$ direction was set to have a constant absorption and scattering parameters of $a=0.01$ m${}^{-1}$ and $b_0=0.3$ m${}^{-1}$ respectively.
The other half of the bulk was set to have a constant absorption of $a=0.01$ m${}^{-1}$, but the scattering parameter $b_1$ was scanned over values of $0.1,0.2,0.3,0.4$ and $0.5$ m${}^{-1}$.

The initial vertex was constrained to the origin, while the final vertex was constrained to a sphere of radius 15 cm at either $\vec{\eta}_0 = (0,0,120)$ or $\vec{\eta}_1 = (60, 7, 104)$ meters, simulating two detectors.
Each detector has its own associated path, and both paths are updated at each iteration.
The conditional detection probability (equation~\ref{eq:detection_prob}) was chosen to be an analytic distribution that approximates the IceCube digital optical module detection efficiency \cite{AartsenMeasurementSouthPole2013}.

To evaluate the convergence of the results, four chains are run in parallel, each taking an equal share of the total number of samples.
The potential scale reduction factor, $\hat{R}$ \cite{gelman1992}, is then computed for these four chains.
As the coordinate space is trans-dimensional, $\hat{R}$ cannot be computed directly on the coordinates.
Instead, it can be computed on overall observables of the chain.
For this demonstration, the total length of the path was chosen.

A single path generated from the ray tracer is used to seed each chain.
As the ray tracer does not produce completely unbiased samples, a burn-in period of 10\% of the total samples was used.

\subsection{Comparison of timing distributions}

\begin{figure}
\includegraphics[width=\linewidth]{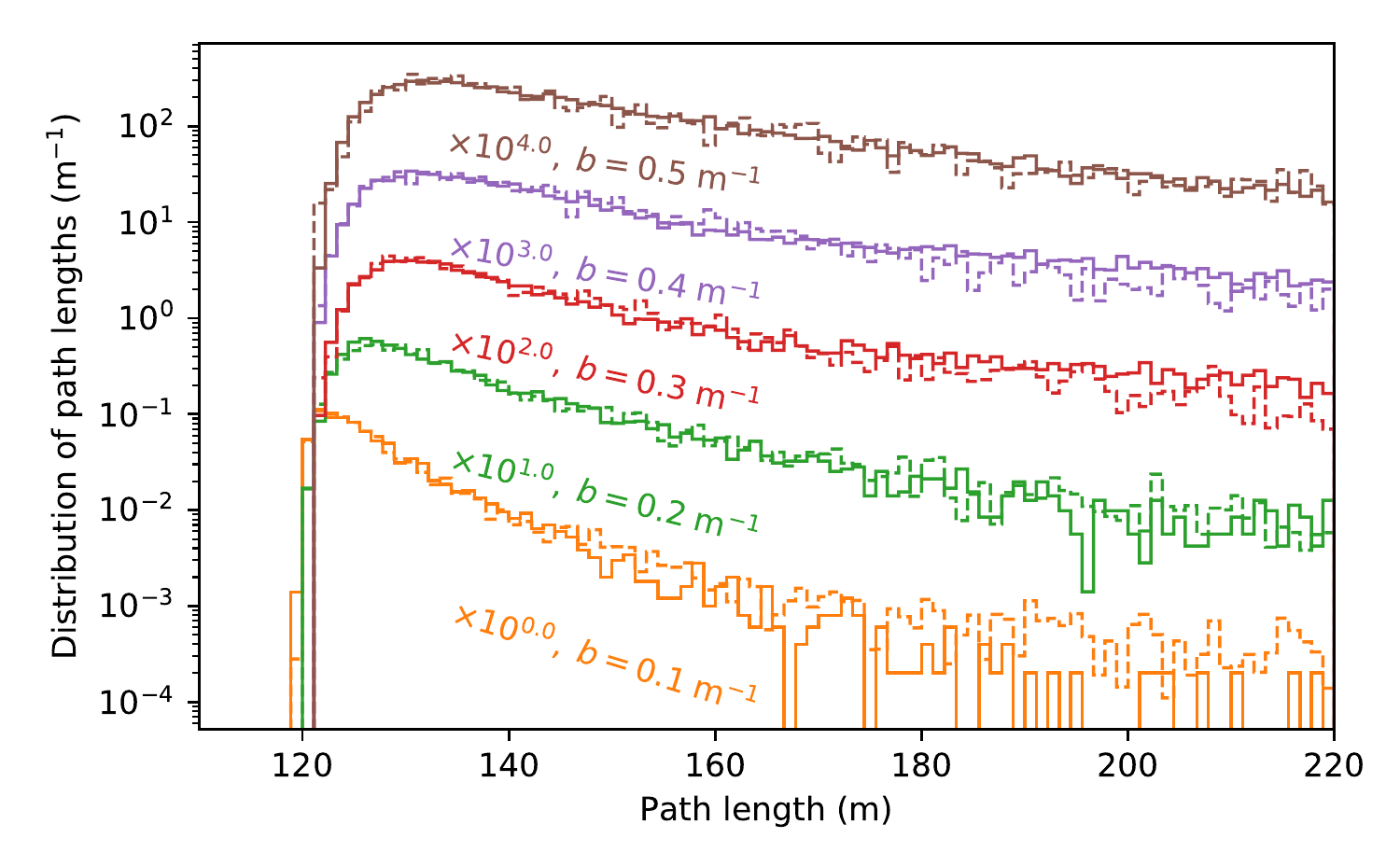}
\caption{\label{fig:time} The distribution of paths lengths for paths that end on detector 0, and various values of the scattering parameter. The solid and dashed lines are generated by the path sampler and ray tracer respectively. Each pair of distributions are offset by the stated value for visual presentation.}
\end{figure}

\begin{table}
\begin{tabular}{c||c|c|c|c|c}
$b_1$ (m${}^{-1}$) & 0.1 & 0.2 & 0.3 & 0.4 & 0.5 \\ \hline \hline
 & \multicolumn{5}{c}{Ray tracer} \\ \hline
CPU Time (s) & 45.9k & 78.1k & 99.6k & 122k & 156k \\ \hline
$\mathcal{B}$ & 0.67(3) & 0.79(2) & 0.73(2) & 0.68(2) & 0.59(2) \\ \hline \hline
 & \multicolumn{5}{c}{Path sampler} \\ \hline
$N_{\text{samples}}$ & 1M & 1.5M & 3.2M & 4.5M & 4.5M \\ \hline
CPU Time (s) & 23.0 & 74.4 & 232 & 373 & 416 \\ \hline
$\mathcal{B}$ & 0.64(7) & 0.77(1) & 0.76(7) & 0.64(4) & 0.50(6) \\ \hline
$\hat{R}$ & 1.11 & 1.08 & 1.19 & 1.12 & 1.02 \\ \hline
$\bar{\alpha}$ & 30\% & 23\% & 20\% & 20\% & 19\% 
\end{tabular}
\caption{\label{tbl:results} The results of five runs of both the ray tracer and path sampler, using different values of the scattering parameter $b_1$ as defined in Figure~\ref{fig:setup}.}
\end{table}

The total amount of light received by a light sensitive element is related to the total amount of absorption and scattering.
The relative amount of scattering to absorption can be measured using the time delay of the light reaching the elements; more scattering leads to longer, meandering paths through the bulk.
The time delay of each ray is directly proportional to the length of the path it takes, as the bulk is assumed to have a constant index of refraction.

Figure~\ref{fig:time} shows a comparison between the distribution of path lengths generated by both the ray tracer and path sampler, for paths than end on detector 0.
Paths that end on detector 1 are shown in Figure~\ref{fig:time2}.
The solid and dashed lines are generated by the path sampler and ray tracer respectively.
The ray tracer was run until 5000 rays were collected; generally the minimum required to provide a good constraint on the scattering parameter \cite{dimapri}.
The path sampler was run until a $\hat{R}$ of less than $1.2$ was achieved, with the results shown in Table~\ref{tbl:results}.

\subsection{Comparison of relative light yield}

The number of photons that reach a detector is proportional to the integral over all paths, and is used to constrain the absorption.
Rather than computing the integral directly, it is easier to find the ratios of these integrals for two detector elements.
This problem is equivalent to finding the Bayes factor in Bayesian inference.
The geometric estimator \cite{Meng96simulatingratios}
\begin{equation}
	\mathcal{B} = \frac{\mathbb{E}_0[ \sqrt{p_1(x)/p_0(x)}]}{\mathbb{E}_1[ \sqrt{p_0(x)/p_1(x)}]}
\end{equation}
is used, where $p_0$ and $p_1$ are the probability distributions under comparison, $\mathbb{E}_0$ is the expectation taken over the paths that end on detector 0, and similarly for $\mathbb{E}_1$. 
Then, $\mathcal{B}$ is the ratio of the amount of light that detector 1 sees to the light yield for detector 0.
Intuitively, the estimator works by comparing the ratio of probabilities for paths that end on one detector or the other. 
The application of this estimator is described in more detail in the appendix.

The estimated relative light yield, $\mathcal{B}$, is shown in Table~\ref{tbl:results}.
The uncertainty on $\mathcal{B}$ is the sample standard deviation of $\mathcal{B}$ over the four chains in the case of the path sampler, and four evenly divided subsets of the rays for the ray tracer.

\section{Discussion}

In the synthetic test, a performance improvement of 300 to 1000 times was observed on a CPU implementation, using an Intel i7-5820K.
The distribution of path lengths produced by the ray tracer and path sampler match to within the statistical uncertainty, with only a slight disagreement visible in the tails of the distribution.
The relative light yields match to within the uncertainty of both methods, and the path sampler is able to achieve $\sim 10\%$ error.
Larger numbers of samples are observed to decrease this error further.
The average acceptance rate, $\bar{\alpha}$ is maintained at $> 10\%$ even in the high scattering scenario.
While this is a highly encouraging result, some caution must be exercised in predicting the performance increase when applied to light simulation in an experiment.

Ray tracing is performed most efficiently on a GPU, where improvements of up to 100 times \cite{AartsenMeasurementSouthPole2013} are possible as ray tracing is highly parallelizable. 
In contrast, an MCMC is an intrinsically serial algorithm.
Nevertheless, multiple chains could be run in parallel; although, the utilization of a GPU will likely be less efficient compared to the ray tracer.

Experiments that use a single -- possibly segmented -- detector will see the most benefit, as the space of all paths under consideration will be contiguous.
Experiments such as IceCube can have thousands of discrete detector elements, requiring mapping functions to connect each subspace of paths.
For the relative light yield between elements, the MCMC must be run until it converges on each element.
In comparison, the ray tracer can be run for less time if an accurate result is not required.

The current implementation uses a simple Metropolis-Hastings style step move using normal distributions.
More advanced methods -- such as Hamiltonian Monte-Carlo methods -- may improve the convergence speed of the MCMC, as they are particularly well suited to problems with hundreds of dimensions.
Any improvement in the convergence of the sampler translates directly into speed gains, as less samples need to be taken.

Scattering can be wavelength dependent, an effect that is not incorporated in this demonstration.
The implementation described can be extended through adding a wavelength parameter to the probability distribution, and sampling over it along with the path vertices.

\section{Conclusion}

The described path sampler reproduces the timing distribution for light traversing a distance of 120 meters in a medium with various scattering parameters.
It also reproduces the relative light yield for two detectors.
A performance improvement of 300 to 1000 times was observed, when compared to the CPU based ray tracer.

Path sampling can be extended to other kinds of experiments.
In smaller neutrino detectors, light may be reflected or refracted from surfaces.
These processes can be handled by combining the formulation presented here with those already developed in Metropolis Light Transport for reflection and refraction.
The initial and final states need not be constrained only in position, they can alternatively be constrained in angle, such as when indirectly imaging light from a star.

Particles other than photons can also be simulated.
General particle physics simulations use ray tracing to propagate particles such as neutrons or muons through matter, and could benefit from path sampling if the final state is constrained.
An example of constrained initial and final states are the simulation of atmospheric showers of particles induced by cosmic rays, where the particle must start as a particular species and end in a detector.
This technique shows promise for simulation of any particles where the initial and final states of the path are highly constrained.

\begin{acknowledgments}
The author would like to thank C.A. Arg\"uelles, D. Chirkin, J. Conrad, J. Heyer, and B.J.P. Jones for their insight and contribution to this work.
The author was supported by NSF-PHY-1505858 and the American-Australian Association Bechtel fellowship.
\end{acknowledgments}

\bibliographystyle{apsrev4-1} 
\bibliography{bib} 

\begin{appendix}
\section{Appendix}

This appendix provides additional detail on the construction of the probability distribution, the reversible jump proposal, the coordinate system in which paths are sampled, the method by which the relative light yield is estimated, and the ray tracer used in the demonstration.

\subsection{Path probability distribution}

\begin{figure}
\includegraphics[width=\linewidth]{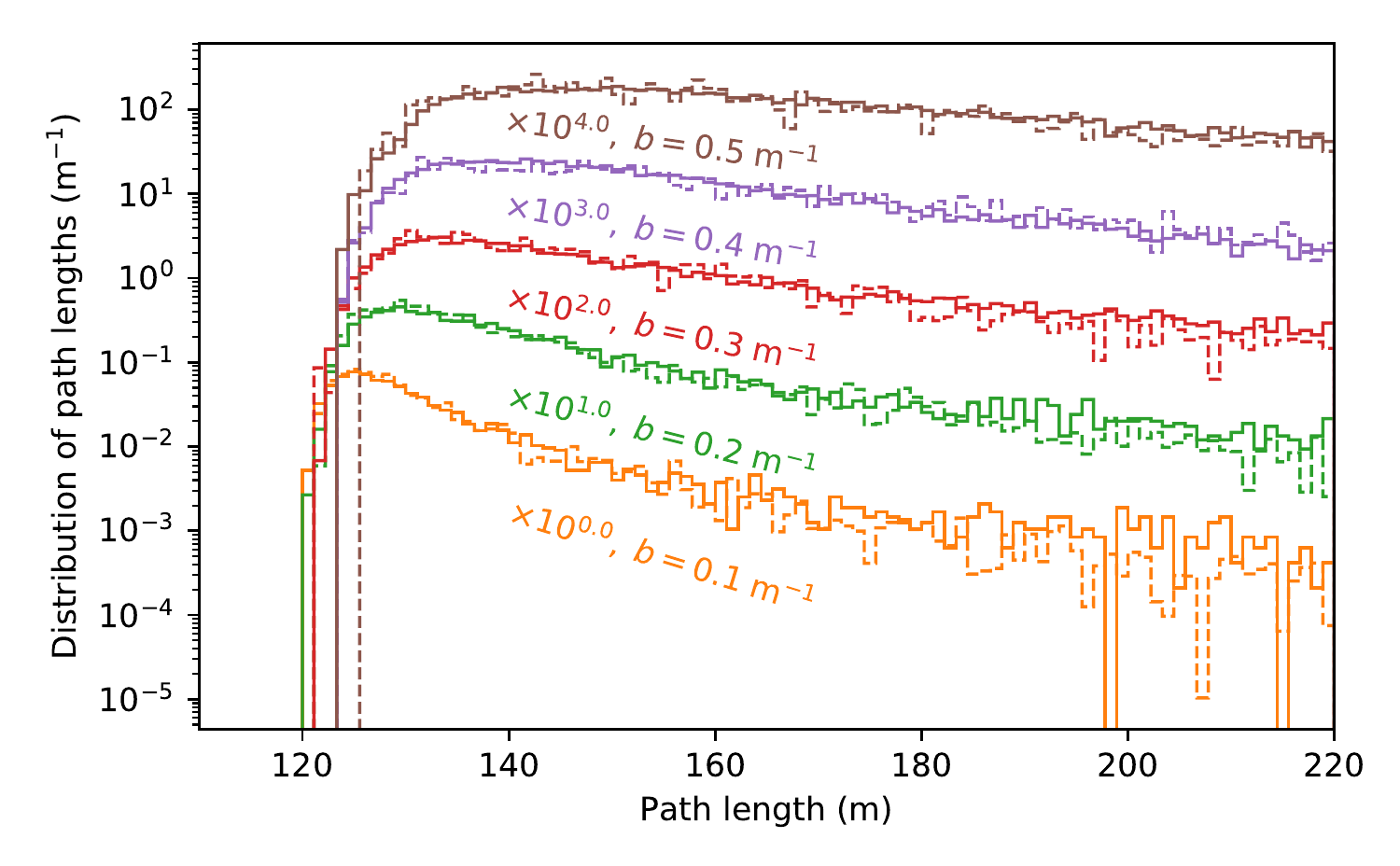}
\caption{\label{fig:time2} The distribution of paths lengths for paths that end on detector 1, and various values of the scattering parameter. The solid and dashed lines are generated by the path sampler and ray tracer respectively. Each pair of distributions are offset by the stated value for visual presentation.}
\end{figure}

\begin{figure*}
\includegraphics[page=4]{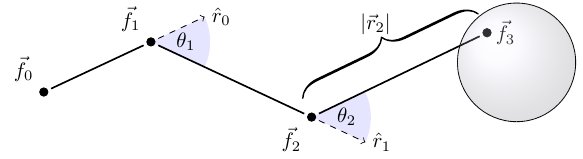}
\caption{\label{fig:path} An example of a path specified as a series of four vertices connected by straight line segments.}
\end{figure*}

It may seem most natural to express the path probability distribution in spherical coordinates.
However, using the example shown in Figure~\ref{fig:path}, the position of vertex $\vec{f}_2$ depends not just on $|\vec{r}_1|$ and $\theta_1$, but also on $|\vec{r}_0|$ and $\theta_0$.
As discussed in the main text, the proposal of new paths takes place in bi-spherical coordinates.
However, these coordinates do not admit a simple construction for the path probability distribution.
Instead, the distribution will be specified in Cartesian coordinates, and a transformation between the two coordinate systems will then be defined.
By using Cartesian coordinates, the probability distribution can be factorized into distributions for the initial (starting) vertex, the intermediate vertices, and the final vertex:
\begin{multline}
	p(\vec{f}_0, \vec{f}_1, \ldots, \vec{f}_n) = p_i(\vec{f}_0, \vec{f}_1) \\ \left[ \prod_{k=1}^{n-2} p_v(\vec{f}_{k-1}, \vec{f}_{k}, \vec{f}_{k+1}) \right] p_f(\vec{f}_{n-2}, \vec{f}_{n-1})
\end{multline}

\subsubsection{Initial vertex}
The initial vertex factor captures the profile of emitted light.
In this derivation, a point source is assumed, such that the first vertex of the path is a fixed location.
This can also be extended to line, surface, or volumetric sources of light.

The pair of 3D coordinates $\vec{f}_0$ and $\vec{f}_1$ give the outgoing direction from the point source located at $\vec{f}_0$. 
The initial vertex probability is
\begin{equation}
	p_i(\vec{f}_0, \vec{f}_1) = \varepsilon(\hat{r}_{0}),
\end{equation}
where $\varepsilon$ is the angular emission distribution.

The angular emission distribution is taken to be a von Mises-Fisher distribution:
\begin{equation}
\varepsilon(\hat{r}_0) = \frac{\kappa e^{\kappa \hat{r}_0\cdot\hat{\varepsilon}}}{4 \pi \sinh{\kappa}}. \label{eq:emission_dist}
\end{equation}
where $\hat{\varepsilon}$ is the preferred direction of the source.
For the demonstration in the main text, this was chosen to be $\hat{\varepsilon}=(0, 0, 1)$.

\subsubsection{Intermediate vertex}
The intermediate vertex distribution is composed of two factors: the probability of the light traveling from the previous vertex $\vec{f}_{k-1}$ to this vertex along a straight line path
\begin{equation}
	\vec{v}_k(s) = \vec{f}_{k-1} + s \vec{r}_{k-1},
\end{equation}
where $\vec{r}_{k-1} = \vec{f}_k - \vec{f}_{k-1}$, then the light scattering at $\vec{f}_k$; and then, the probability of the light changing direction.
The probability of traveling this distance is given by the product of two distributions: the probability density function of the ray scattering at $\vec{f}_k$, and the cumulative distribution function of the ray not being absorbed along this journey, of which both distributions are exponential.
The probability of changing direction is given by the angular scattering distribution $\sigma$.
Thus, the intermediate vertex probability is
\begin{equation}
	p_v(\vec{f}_{k-1}, \vec{f}_k, \vec{f}_{k+1}) = \frac{ b(\vec{f}_k) e^{- \tau(k)} }{ |\vec{r}_{k-1}|^2 } \sigma(\cos \theta_{k}), \label{intervtx}
\end{equation}
where $\cos \theta_{k} = \hat{r}_{k-1} \cdot \hat{r}_{k}$, and $\tau(k)$ is the absorption and scattering depth:
\begin{align}
	\tau(k) &= \tau_a(k) + \tau_b(k), \\ 
	\tau_a(k) &= \int_{0}^{1} a(\vec{v}_{k}(s)) ds, \\ 
	\tau_b(k) &= \int_{0}^{1} b(\vec{v}_{k}(s)) ds. \label{eq:b_depth_def}
\end{align}

Following the parameterization of \textcite{AartsenMeasurementSouthPole2013}, a mixture of the simplified Liu and Henyey-Greenstein distributions were used for the angular scattering distribution:
\begin{equation}
	\sigma(\cos \theta) = f_{\textrm{SL}} p_{\textrm{SL}}(\cos \theta) + (1-f_{\textrm{SL}}) p_{\textrm{HG}}(\cos \theta) \label{eq:angular_scattering}
\end{equation}
where
\begin{equation}
	p_{\textrm{SL}}(\cos \theta) = \frac{1}{2} \left( \frac{1+\Gamma}{1-\Gamma} \right) \left( \frac{1 + \cos{\theta}}{2} \right)^{2 (\Gamma^{-1}-1)^{-1}}
\end{equation}
is the simplified Liu distribution,
\begin{equation}
	p_{\textrm{HG}}(\cos \theta) = \frac{1}{2} \left(1 - \Gamma^2\right) \left( 1 + \Gamma^2 - 2 \Gamma \cos{\theta} \right)^{-3/2}
\end{equation}
is the Henyey-Greenstein distribution, and $\Gamma=\langle \cos \theta \rangle$ is the average $\cos \theta$ of the distribution. 
In the demonstration of the main text, the values of $\Gamma = 0.95$ and $f_{\textrm{SL}} = 0.45$ were used.

\subsubsection{Final vertex}
Light sensitive elements generally have a 2D topology, and in this derivation a spherical element is assumed.
The light ray must travel the final segment without scattering or absorption, so the probability of each is given by a c.d.f.\ --- in contrast to the p.d.f.\ used for scattering in the intermediate vertex factor.
The constraint that the final vertex must be located on the element surface introduces a $\cos{\Phi}/r^2$ geometric term, where $\Phi$ is the angle of the incoming ray to the surface normal.
The final vertex probability is
\begin{equation}
	p_f(\vec{f}_{n-2}, \vec{f}_{n-1}) = \frac{\max(0, - \hat{r}_{n-2} \cdot \vec{\nu}(\vec{f}_{n-1})) }{ |\vec{r}_{n-2}|^2 } \rho(\vec{f}_{n-1}),
\end{equation}
where $\vec{\nu}(\vec{f}_{n-1})$ is the surface normal at $\vec{f}_{n-1}$, and $\rho$ is the probability of detection conditioned on the vertex location $\vec{f}_{n-1}$.

The functional form of $\rho$ was chosen so that it would roughly match the detector response shown in \textcite{AartsenMeasurementSouthPole2013}:
\begin{equation}
	\rho(\vec{f}) = \frac{\exp(3 \vec{\nu}(\vec{f}) \cdot \hat{\rho} - 1)}{\cosh(2 \vec{\nu}(\vec{f}) \cdot \hat{\rho} + 0.7)} \label{eq:detection_prob}
\end{equation}
where $\hat{\rho}$ is the preferred direction of the detector.
For the demonstration in the main text, this was chosen to be $\hat{\rho} = (0, -1, 0)$.

\subsection{Jump proposal}
A pair of adjacent vertices ($k$, and $k+1$) are chosen with probability proportional to the scattering depth, $\tau_b(k)$ between them:
\begin{equation}
	p(n \rightarrow n+1, k) = \frac{\tau_b(k)}{\sum_{l=1}^{n-1} \tau_b(l) },
\end{equation}
There is equal probability of removing any intermediate vertex:
\begin{equation}
	p(n \rightarrow n-1, k) = \frac{1}{n-2}.
\end{equation}

\subsection{Dimension matching transformation}

Figure~\ref{fig:bisph} shows that the new vertex $\vec{f}'$ is placed between the existing vertices $\vec{f}_k$ and $\vec{f}_{k+1}$ which form the foci for the coordinate system.
\begin{equation}
	\vec{f}'(s,t,\phi) = \vec{f}_k + \gamma \frac{\hat{r}_k \sinh{t} + ( \hat{\xi} \cos{\phi} + \hat{\zeta} \sin{\phi} )\sin{s} }{\cosh{t} - \cos{s}},
\end{equation}
where both $\hat{\xi}$ and $\hat{\zeta}$ are mutually perpendicular unit vectors to $\hat{r}_{k}$, and $\gamma = |\vec{r}_{k}|/2$.
The Jacobian factor for this transformation is
\begin{equation}
	J(s, t, \phi) = \left( \frac{\gamma}{\cosh{t} - \cos{s}} \right)^3 \sin{s},
\end{equation}
and the $q(s,t,\phi)$ distribution is factorable into three independent distributions:
\begin{align}
	q(s) &= \frac{\beta e^{-\beta \cos{s} }}{ 2  \sinh{\beta} } \sin{s}, \\
	q(t) &= (2 + 2\cosh{t})^{-1}, \\
	q(\phi) &= \frac{1}{2\pi},
\end{align}
where $\beta$ is a tunable parameter.

The probability of accepting a jump proposed with $\vec{f'}(s,t,\phi)$ and $q(s)q(t)q(\phi)$ is maximized when the curvature of
\begin{equation}
	\frac{p_v(\vec{f}_k, \vec{f}'(s,t,\phi), \vec{f}_{k+1})}{q(s) q(t) q(\phi)} J(s, t, \phi)
\end{equation}
is unity.
Matching the curvature of $q$ and $J$ to $p$ is known as centering the distribution \cite{FanReversiblejumpMarkov2010}.
In this case, an exact match is not possible due to the effects of the adjacent vertices.
A partial match can be achieved by setting
\begin{equation}
	\beta = \kappa_{\textrm{eq}} + \frac{\gamma b(\vec{f'}) - 1}{2}.
\end{equation}
where $\kappa_{\textrm{eq}}$ is the concentration parameter $\kappa$ for a von Mises-Fisher distribution for which the average $\cos \theta$ is equal to the $\Gamma$ of the angular scattering distribution:
\begin{equation}
	\coth{\kappa_{\textrm{eq}}} - \frac{1}{\kappa_{\textrm{eq}}} = \Gamma.
\end{equation}

\subsection{Path space coordinate system}

\begin{figure*}
\includegraphics[page=5]{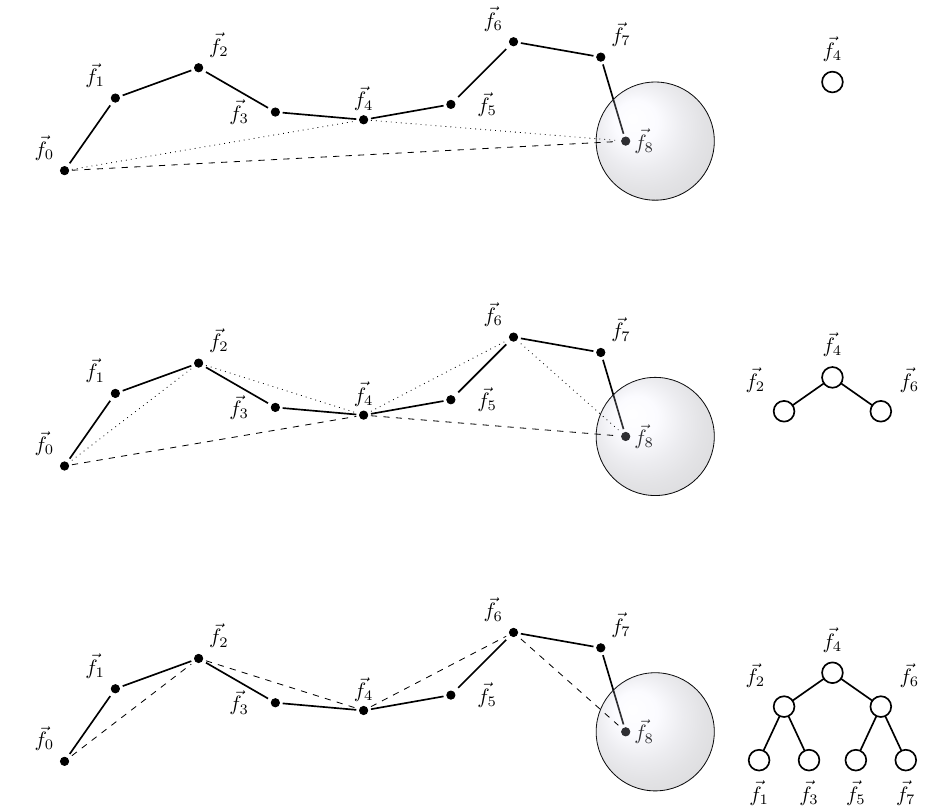}
\caption{\label{fig:tree_coords} The recursive structure in which the coordinates for sampling are defined. \textbf{Top}: $\vec{f}_0$ and $\vec{f}_8$ define the foci of a bi-spherical coordinate system in which the location of $\vec{f}_4$ is defined. \textbf{Middle}: $\vec{f}_0$ and $\vec{f}_4$ are used as foci to define $\vec{f}_2$. In addition, $\vec{f}_4$ and $\vec{f}_8$ are used as foci to define $\vec{f}_6$. This forms a tree-like dependency structure between $\vec{f}_2$, $\vec{f}_6$ and $\vec{f}_4$. \textbf{Bottom}: The process is repeated, and $\vec{f}_1,\vec{f}_3,\vec{f}_5,$ and $\vec{f}_7$ become children of $\vec{f}_2$ and $\vec{f}_6$. The dashed lines show the primary axis of the bi-spherical coordinate system. The dotted lines connect the foci to the vertex being defined, and then become the primary axes of the next iteration.}
\end{figure*}

\begin{figure*}
\includegraphics[page=6]{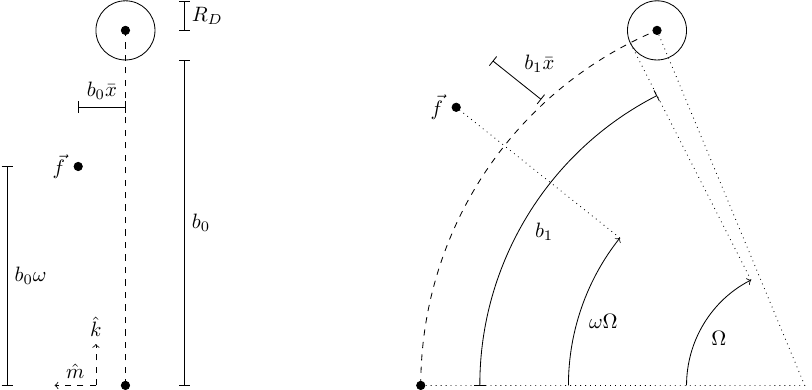}
\caption{\label{fig:endxform} A vertex, $\vec{f}$, that belongs to a path that ends on detector 0 (\textbf{left}) is transformed to a path that ends on detector 1 (\textbf{right}). The transformation is defined such that points that lie on the dashed lines map to each other.}
\end{figure*}

The coordinates in which the probability distribution is sampled are defined in terms of a transformation to Cartesian coordinates.

The transformation is defined as follows, and is shown in Figure~\ref{fig:tree_coords}.
Given two coordinates in Cartesian space, $\vec{f}_k$ and $\vec{f}_l$, a bi-spherical coordinate system is set up with these coordinates as the foci.
The vertex mid-way between these two, with index $m = \lfloor (k+l)/2 \rfloor$, will now be defined in terms of the bi-spherical coordinates $s_i, t_i, \phi_i$.
Once $\vec{f}_m$ has been found in Cartesian coordinates, it can be used as a focus to define two other vertices.
Thus, this process is repeated twice: once with foci $\vec{f}_k$ and $\vec{f}_m$, and then again with foci $\vec{f}_m$ and $\vec{f}_l$.

It should be noted that, while the vector $\vec{f}_{k\rightarrow l} = \vec{f}_l - \vec{f}_k$ is used to define one axis of the bi-spherical coordinate system, two other axis are needed.
These axes are inherited by the parent iteration of the recursive definition.
Then, the axes are rotated by $\phi_i$ around the plane of the axes, then rotated by the angle between $\hat{f}_{k \rightarrow l}$ and $\vec{f}_{k \rightarrow m}$, then passed to the child iteration that uses $\vec{f}_k$ and $\vec{f}_m$ as the foci. 
The same process is repeated for passing the axes to the child iteration that uses $\vec{f}_m$ and $\vec{f}_l$.
In this way, a rotation of one vertex also rotates all child vertices that depend on it.

This recursive definition is started with the initial and final vertex $\vec{f}_0$ and $\vec{f}_{n-1}$ as the foci. 
It ends when $l=k+1$, i.e.\ when there are no remaining vertices in the path.
The definition allows large scale movements of the path, without the lever arm effect that is present in a spherical coordinate system.

Sampling directly in $s_i$ space can be problematic, as changes in $s$ when $s$ is near zero cause large changes in the location of the vertex in Cartesian coordinates.
This problem can be partially alleviated through another change in coordinates, this time using $\zeta_i$ instead of $s_i$ where
\begin{equation}
	\zeta_i = \tanh^{-1}{(\cos s_i)}.
\end{equation}
This new coordinate has the advantage of being unbounded, whereas $s_i$ must lie in the interval of $[0, \pi]$.
As $\zeta_i$ becomes large or small, it approaches an $s_i$ of $0$ or $\pi$.

Each iteration of this recursion also contributes a factor to the Jacobian of the transformation, which is given by:
\begin{equation}
	J = \prod_{i} \left ( \frac{\gamma_i}{\cosh{t_i} - \tanh{\zeta_i}} \right)^3 \cosh^2{\zeta_i}
\end{equation}

\subsection{Relative light yield transformation}

As noted in the main text, finding the relative light yield, $\mathcal{B}$, is related to computing the Bayes factor for two models with shared parameters.
Usually, the two models have different probability distributions $p_0$ and $p_1$, but have an equivalent volume of parameter space that the models range over.
In this case, the probability distribution for paths that end on detector 0 is the same as that for detector 1.
Instead, it is the volume of parameter space that is different, as each detector has a different restriction on the location of the last vertex in the path.

To use the same machinery defined for Bayes factors, we can define a mapping between paths that end on one detector and paths that end on the other.
Then the paths that end on detector 1 can be specified in terms of coordinates for paths that end on detector 0, by use of a transformation function $T(\{\vec{f}_i\})$.
If we let the path probability distribution $p$ be the probability distribution $p_0$ for one model, the probability distribution for the other model is modified by a factor of the Jacobian of T:
\begin{equation}
	p_1(\{\vec{f}_i\}) = p_0(T(\{\vec{f}_i\})) \left |\frac{\partial T(\{\vec{f}_i\})}{\partial \{\vec{f}_i\}} \right|
\end{equation}
Now, the relative light yield is specified in terms of two probability distributions $p_0$ and $p_1$ with a shared coordinate space, and the machinery of Bayesian inference can be used.

Ideally, $T$ should be chosen such that paths of high probability that end on detector 0 map to paths of high probability that end on detector 1.
A simple rotation of the path will not suffice, as the emission probability distribution $\varepsilon$ will heavily penalize the sharp turn created at the initial vertex.
Instead, the path is smoothly curved toward detector 1 by deforming it along a cylinder that intersects the initial vertex and detector 1.
The radius of curvature for the cylinder is chosen such that the tangent of the cylinder at the initial vertex matches the preferred direction of the emission distribution, $\hat{\varepsilon}$.

To perform the transformation, the vertices for the path that ends on detector 0 are first written in terms of an intermediate coordinate system, $\bar{x}, \bar{y}, \omega$.
Let
\begin{align}
	b_0 &= | \vec{\eta}_0 - \vec{f}_0 | - R_D, \\
	\hat{k} &= \frac{\vec{\eta}_0 - \vec{f}_0}{| \vec{\eta}_0 - \vec{f}_0 | }
\end{align}
where $\vec{\eta}_0$ is the location of detector 0, and $R_D$ is the radius of the detector.
Let $\hat{n}$ and $\hat{m}$ be two mutually perpendicular unit vectors to $\hat{k}$.
Then, the intermediate coordinates for the vertex with index $i$ is:
\begin{align}
	\bar{x}_i &= \frac{(\vec{f}_i - \vec{f}_0) \cdot \hat{m}}{b_0},& \bar{y}_i &= \frac{(\vec{f}_i - \vec{f}_0) \cdot \hat{n}}{b_0}, \\
	\omega_i &= \frac{(\vec{f}_i - \vec{f}_0) \cdot \hat{k}}{b_0}.
\end{align}

From these, the path that ends on detector 1 is generated as follows.
Let $\vec{B} = \vec{\eta}_1 - \vec{f}_0$, then the major axis of the cylinder will be perpendicular to $\hat{\varepsilon}$ and $\hat{B}$.
The radius of curvature for this cylinder is
\begin{equation}
	r_C = \frac{|\vec{B}|}{2} \frac{1}{\sin \Psi},
\end{equation}
where $cos \Psi = \hat{\varepsilon} \cdot \hat{B}$.
The angle subtended by the initial vertex, and the location where the cylinder intersects the detector, is roughly
\begin{equation}
	\Omega = 2 \Psi - \frac{R_D}{r_C}.
\end{equation}
Thus the distance along the cylinder between the initial vertex and the detector is $b_1 = r_C \Omega$.

Then, the Cartesian coordinates of the vertex with index $i$ is:
\begin{widetext}
\begin{align}
	\vec{f}_i = \left(\vec{f}_0 - r_C \hat{m}\right) + \left(r_C + \bar{x}_i b_1\right) \left( \hat{m} \cos(\omega_i \Omega) + \hat{k} \sin(\omega_i \Omega)\right) + \hat{n} \bar{y}_i b_1
\end{align}
\end{widetext}

The last vertex is included in this transformation; however, the transformation may have left it sitting away from the surface of the detector, so it is projected back onto the surface as a final step.

The Jacobian of this transformation is
\begin{equation}
	J = \prod_{i=1}^{n-2} \frac{b_0^3}{b_1^2 \Omega ( r_C + \bar{x}_i b_1) } 
\end{equation}

\subsection{Ray tracer}

\begin{figure}
\begin{minipage}{\linewidth}
\begin{algorithm}[H]
\caption{Ray tracer}\label{ag:ray_trace}
\begin{algorithmic}[1]
\State $\vec{x} \gets \text{initial location}$.
\State $\vec{n} \gets \text{initial direction}$
\State $\ln(w) \gets 1 \text{ (initial weight)}$
\For {$i \gets 0, \ldots, i_\text{max}$}
	\State $\tau_b \gets -\ln(\textit{uniform}(0, 1))$
	\State $\Delta L \gets \text{solve } \int_0^{\Delta L} b(\vec{x} + s\hat{n}) ds = \tau_b$
	\State $d \gets \text{straight line distance to target } \vec{\eta}$
	\State $\tau_a \gets \int_0^{\min(d, \Delta L)} a(\vec{x} + s \hat{n}) ds$
	\State $\ln(w) \gets \ln(w) - \tau_a$
	\If {$d < \Delta L$}
		\State \text{accept path}	
	\Else
		\If {$| \vec{x} - \vec{\eta} | > 200 \text{ m}$}
			\State \text{reject path}
		\Else
			\State $\vec{x} \gets \vec{x} + \Delta L \hat{n}$
			\State $\cos{\theta} \gets \text{sample from }\sigma(\cos{\theta})$
			\State $\phi \gets \textit{uniform}(0, 2\pi)$
			\State $\hat{n} \gets \textit{rotate}(\hat{n}, \cos{\theta}, \phi)$
		\EndIf
	\EndIf
\EndFor
\end{algorithmic}
\end{algorithm}
\end{minipage}
\end{figure}

The ray tracer proceeds as follows, also shown in Algorithm~\ref{ag:ray_trace}.
It keeps a current state of the ray, including the current direction and position of the ray.
The initial location of the ray is set to the point-source of light at $(0, 0, 0)$ meters.
The initial direction is sampled from the emission distribution $\varepsilon$.

The ray tracer then advances the ray along the current direction.
The distance to advance is based on the scattering length of the bulk.
Generally, it is the solution $\Delta L$ to 
\begin{equation}
	 \int_0^{\Delta L} b(\vec{x} + s\hat{n}) ds = \tau_b
\end{equation}
where $\tau_b$ has been drawn from an exponential distribution.
For the test case described in Figure~\ref{fig:setup}, the solution to this equation is analytic.
If the direction of the ray intersects with a detector, the straight line distance to the detector along the ray is calculated.
If this distance is less than $\Delta L$, such that advancing the ray would cause it to travel through the detector, the ray is accepted and recorded.
Otherwise, the current position is updated by advancing the ray by $\Delta L$, and a new direction is sampled from the angular scattering distribution.

At each iteration, a sample weight for the ray is also kept.
When the ray advances, the weight is reduced by the probability that the light is not absorbed.
If the ray were to be accepted, the straight line distance to the detector is used to calculate this probability, otherwise $\Delta L$ is used.

At each iteration, the current geometric distance to the detector is calculated.
If this distance exceeds 200 meters, the ray is deemed to have wandered too far from the detector and it is rejected and not recorded.
Once a ray is either accepted or rejected, the algorithm is started again from the beginning for the next ray.

When multiple detectors are involved, the potential intersection and straight line distance of the ray to each is computed.
If the ray intersects any detector, it is accepted and recorded.
The ray is rejected if the distances to each detector are both greater than 200 meters.

\end{appendix}

\end{document}